\documentclass[article,accept,moreauthors,pdftex,10pt,a4paper,preprints]{mdpi} 

\firstpage{1} 

\history{}

\usepackage[utf8]{inputenc}
\usepackage[T1]{fontenc}
\usepackage{hyperref}
\usepackage{graphicx,bm,amsmath,color,scalerel}
\usepackage{bbold}
\usepackage{wasysym}
\usepackage{verbatim}

\newcommand{\be}{\begin{equation}}
\newcommand{\ee}{\end{equation}}
\newcommand{\beq}{\begin{eqnarray}}
\newcommand{\eeq}{\end{eqnarray}}
\newcommand{\ba}{\begin{align}}
\newcommand{\ea}{\end{align}}

\Title{Twin Peaks: A Possible Signal in the Production of Resonances beyond Special Relativity}

\Author{Germán Albalate $^{1}$\orcidA{}, José Manuel Carmona $^{2}$*\orcidB{}, José Luis Cortés $^{2}$\orcidC{} and José Javier Relancio $^{2}$\orcidD{}}

\AuthorNames{Germán Albalate, José Manuel Carmona, José Luis Cortés and José Javier Relancio}

\address{%
$^{1}$ \quad Departamento de F\'{\i}sica de la Materia Condensada, Universidad de Zaragoza, Zaragoza 50009, Spain; galbalate@unizar.es\\
$^{2}$ \quad Departamento de F\'{\i}sica Te\'orica,
Universidad de Zaragoza, Zaragoza 50009, Spain; jcarmona@unizar.es (J.M.C); cortes@unizar.es (J.L.C.); relancio@unizar.es (J.J.R.)}

\corres{Correspondence: jcarmona@unizar.es}

\abstract{It is usually expected that quantum gravity corrections will modify somehow the symmetries of special relativity. In this paper we point out that the possibility of very low-energy (with respect to the Planck energy) modifications to special relativity in the framework of a deformed relativistic theory is not ruled out, and that, depending on the value of that scale, such a possibility could be tested in accelerator physics. In particular, we take a simple example of a relativistic kinematics beyond special relativity from the literature, and obtain a remarkable effect: two correlated peaks \emph{(twin peaks)} associated to a single resonance. We analyze this phenomenology in detail, use LEP data to put constraints of the order of TeV on the scale of corrections to special relativity, and note that such an effect might be observable in a future very high-energy proton collider.}


\begin{document}

\section{Introduction}

Special relativity (SR) is a fundamental ingredient of quantum field theory (QFT), which constitutes the theoretical framework for the description of interactions in particle physics. However, it is well known that when one wants to put together gravity with QFT, one finds non-renormalizable infinities. In particular, Feynman showed~\cite{Feynman:1996kb} that if one considers an interaction mediated by a spin 2 particle, one obtains the same equations as in general relativity (GR), where the gravitational interaction is introduced by going from a quantum field theory in flat spacetime to curved spacetime (see Ref.~\cite{Birrell:1982ix} for the formulation of quantum field theory in curved spacetime, and also Ref.~\cite{Bogoliubov:1982book} for the Bogoliubov transformations relating the Fock spaces of accelerated and inertial observers). Such theory, however, turns out to be non-renormalizable for energies comparable with the Planck scale.

There have been many attempts to avoid the problems of inconsistency between GR and QFT, including string theory~\cite{Mukhi:2011zz,Aharony:1999ks,Dienes:1996du}, loop quantum gravity~\cite{Sahlmann:2010zf,Dupuis:2012yw}, supergravity~\cite{VanNieuwenhuizen:1981ae,Taylor:1983su}, or causal set theory~\cite{Wallden:2013kka,Wallden:2010sh,Henson:2006kf}. In most of these theories a minimum length appears~\cite{Gross:1987ar,Amati:1988tn,Garay1995}, which is normally associated to the Planck length $\ell_P \sim 1.6\times 10^{-33}$\,cm. Qualitatively speaking, at such a distance quantum effects should replace the continuum spacetime by some sort of ``space-time foam''~\cite{Wheeler:1955zz,Ng:2011rn}. While we do not fully understand yet this regime, the previous arguments suggest that the symmetry of a continuum spacetime, that is, Poincaré invariance, is only a low-energy symmetry, so that it seems reasonable to expect that special relativity will be modified at a certain energy scale by the new physics. It is then interesting to try to figure out the phenomenological windows where these modifications could be perceived. In Feynman's words~\cite{Feynman:1963uxa}: \emph{Today we say that the law of relativity is supposed to be true at all energies, but someday somebody may come along and say how stupid we were. We do not know where we are ``stupid'' until we ``stick our neck out,'' and so the whole idea is to put our neck out. And the only way to find out that we are wrong is to find out \emph{what} our predictions are. It is absolutely necessary to make constructs.}

Signals from a modification of SR may be envisaged in two different frameworks: as a Lorentz invariance violation (LIV)~\cite{Kostelecky:2008ts}, in which there is a privileged system of reference, or as a deformation of special relativity (DSR)~\cite{AmelinoCamelia:2008qg}, where there is still a relativity principle. The possible phenomenology and constraints~\cite{Mattingly:2005re,Liberati2013} are quite different in these two cases: while LIV includes very sensitive effects, such as large modifications in thresholds of reactions or energy-loss mechanisms through decay channels which are forbidden in SR, the existence of a relativity principle suppresses or even inhibits such effects for DSR, leaving time of flight studies as the only relevant phenomenological window identified up to now in the DSR case~\cite{Mattingly:2005re,Carmona:2018xwm}. 

Thinking of SR as a low-energy symmetry of Nature, a natural way to modify it is through corrections parametrized by a high-energy scale $\Lambda$. Quantum gravity arguments suggest that $\Lambda$ would not be far from the energy scale associated to the Planck length, the Planck energy (or mass; we will work in natural units, $c=\hbar=1$) $E_P\approx 1.2\times 10^{19}$\,GeV. In this case, astrophysics at very high energies would be the best suited place to look for such corrections~\cite{Mattingly:2005re,Liberati2013}.

However, if indeed SR is valid only at low energies, one can ask whether modifications to SR could start much earlier than at the Planck scale. In fact, this would be a natural scenario in theories with large extra dimensions, in which the Planck energy is an effective four-dimensional scale, whereas the fundamental scale of the gravitational interaction might be just a few orders of magnitude above the electroweak scale~\cite{ArkaniHamed:1998rs,Antoniadis:1998ig,Sundrum:1998ns,ArkaniHamed:1998kx}.
While there have been attempts to observe a number of consequences from the existence of extra dimensions in LHC results (for a review, see the \emph{Extra Dimensions Searches} section in Ref.~\cite{PhysRevD.98.030001}), little attention has been paid up to now to the possibility of observing modifications to SR in accelerator physics. The main reason is the strong constraints for LIV coming from a number of precision experiments and astrophysical observations (see, e.g.,~\cite{Mattingly:2005re,Liberati2013}, and the recent work~\cite{Aartsen:2017ibm} exploring possible violations to SR in neutrino physics). 

These strong bounds, which apply to the LIV scenario and in some cases put the scale $\Lambda$ close to, or even exceeding, the Planck scale, also include the analysis of possible photon time delays~\cite{Albert:2007qk,Martinez:2008ki,Ackermann:2009aa,HESS:2011aa,Nemiroff:2011fk,Vasileiou:2013vra,Vasileiou:2015wja}. However, there is a recent discussion on the consistency of such constraints in DSR models~\cite{Carmona:2017oit,Carmona:2018xwm}, which rises the possibility to have scenarios compatible with a relativity principle (and therefore immune to most of LIV limits) which do not contain photon time delays. This means that the existence of a relativistic generalization of SR driven by a mass scale $\Lambda$ many orders of magnitude below the Planck mass is not phenomenologically absurd, and opens up the opportunity to test it in high-energy particle physics experiments.

In the present work, we will work out a simplified approach to such an analysis and consider the possible signals in the production of a resonance by making use of a DSR-based model for the modifications to the SR formula of the Breit-Wigner distribution. We will see that a remarkable possible outcome is the apparition of a correlated double peak which is associated to a single resonance, a phenomenon that we have dubbed \emph{Twin Peaks}. The theoretical analysis of this effect will be the objective of Sec.~\ref{sec:twin-peaks}, where some technical details are left for the Appendix. Then, on Sec.~\ref{sec:limits} we will apply the previous result to the case of a resonance in the scattering of two particles, taking $Z$ production at LEP as the prominent example, and to the case of multi-scattering resonance production in a hadron collider, extracting a lower limit estimate of the scale $\Lambda$ and considering the prospects for future searches at a very high energy proton-proton collider. 

Sections~\ref{sec:twin-peaks} and~\ref{sec:limits} contain the main message and results of the paper, which are based on the qualitative model for the Breit-Wigner formula in a relativistic extension of SR. We have included however a more careful analysis in Sec.~\ref{sec:cross-section} of the total cross section for the simple process $f_i \overline{f}_i \to X \to f_j \overline{f}_j$ (two-particle production and decay of a resonance $X$), which essentially confirms the model used in Sec.~\ref{sec:twin-peaks}. Finally, we will discuss the results and conclude in Sec.~\ref{sec:conclusions}.  

\section{Twin Peaks}
\label{sec:twin-peaks}

As explained in the Introduction, our objective in this section will be to consider a DSR-inspired model for deviations of SR at an energy scale $\Lambda$ which could be much smaller than the Planck energy, $\Lambda\ll M_P$, and try to identify a possible signal in the production of a resonance at a particle accelerator. We will indeed see that the production of a new resonance has unexpected signals if the mass of the particle is of the order of this scale.

Our departure point to introduce the corrections produced by a deformed relativistic kinematics will be the standard relativistic expression of the Breit-Wigner distribution
\be
f(q^2)=\frac{K}{(q^2-M^2_X)^2+M_X^2\Gamma_X^2},
\label{eq:BW}
\ee
where $q^2$ is the squared of the four-momentum of the resonance $X$, $M_X$ and $\Gamma_X$ are its mass and decay width, respectively, and $K$ is a kinematic factor that can be taken approximately constant in the region $q^2\sim M_X^2$ (that is, $K$ is a smooth function of $q^2$ near $M_X^2$).

For a resonance produced by the scattering of two particles, or which decays into two particles, $q^2$ will be the squared of the invariant mass of the two-particle system. In SR, the squared of the invariant mass for a two-particle system with four-momenta $p$ and $\overline{p}$ is
\be
\begin{split}
m^2&=(p+\overline{p})_\mu (p+\overline{p})^\mu\\
&=(p+\overline{p})_0^2-\sum_i(p+\overline{p})_i^2\\
&=E^2+\overline{E}^2+2E\overline{E}-\sum_i p_i^2-\sum_i \overline{p}_i^2-2p\overline{p}\cos\theta \approx 2E\overline{E}(1-\cos\theta),
\end{split}
\label{eq:s2}
\ee
with $\theta$ the angle formed between the directions of the particles, and the last expression appears in the ultra-relativistic limit $(E\sim p)$. 

In DSR theories, the kinematics of SR is modified by, in general, a deformation of the standard relativistic dispersion relation, $E^2=p^2+m^2$, together with a modified composition law (MCL) for the energy and momentum of a system of particles. The necessity to incorporate a MCL as an ingredient of the generalized kinematics is in fact the main characteristic feature of DSR, in contrast to the LIV approach. The reason is that 
the relativity principle, through the existence of new non-linear deformed Lorentz transformations in DSR, relates the modifications in the dispersion relation and in the composition law~\cite{Carmona2012,Carmona2016b}. This is also the source of the differences in the phenomenology of LIV and DSR models~\cite{Carmona:2018xwm}. 

Our \emph{ansatz} will be to maintain the form of the Breit-Wigner distribution, Eq.~\eqref{eq:BW}, in the new kinematics beyond SR (BSR), while modifying the expression of the square of the invariant mass of the two-particle system, Eq.~\eqref{eq:s2}; that is, the deformation of the kinematics will be introduced through a modified relationship between the momentum of the resonance $q$ and the momenta $p$ and $\overline{p}$. This choice amounts to consider a deformed relativistic kinematics in which the dispersion relation is the standard one of SR, while the total momentum of the two-particle system is a non-linear combination of the two momenta of the particles. 

It turns out that it is indeed possible in DSR to have a MCL together with a standard dispersion relation. In fact, from the point of view of Hopf algebras~\cite{Majid1994}, which is the mathematical language of DSR (the MCL is viewed as the \emph{coproduct} in this language~\cite{Carmona:2017cry,Carmona2016b}), one can always make a change of basis in momentum space and work in the so-called \emph{classical basis}~\cite{Borowiec2010,Kowalski-Glikman2002}, in which the Casimir of the algebra (the dispersion relation) is the standard one. A particularly simple case (although it does not correspond to any coproduct of Hopf algebras) in which this situation is realized is when the MCL is \emph{covariant}~\cite{Battisti:2010sr}, that is, invariant under standard (linear) Lorentz transformations. The simplest example is
\begin{equation}\label{eq:MCL}
\mu^2:=(p\oplus\overline{p})^2=m^2\left(1+\epsilon\frac{m^2}{\Lambda^2}\right),
\end{equation}
where $\mu^2$ is the new expression for the squared of the invariant mass of the two-particle system, which modifies the expression in SR, $m^2$, by the dimensionless quantity $\epsilon m^2/\Lambda^2$, where the parameter $\epsilon=\pm 1$ takes into account the two possible signs of the correction.

The expression of the Breit-Wigner distribution, Eq.~\eqref{eq:BW}, as a function of $m^2$, will then change from $f_\text{SR}=f(q^2=m^2)$ to $f_\text{BSR}=f(q^2=\mu^2)$.\footnote{This simple form of the Breit-Wigner distribution is due to the assumption that the dispersion relation is not modified in the example we are considering. In general, the dispersion relation in DSR will be modified and there will be in correspondence a modification of the Breit-Wigner form of the resonance.} In the absence of a full dynamic framework, the fact that a MCL generates such a change in the Breit-Wigner distribution as a function of $m^2$ is at this point an ansatz or an educated guess, although we will propose a specific cross section calculation with such an assumption in Sec.~\ref{sec:cross-section}. We therefore have
\be
\label{eq:BW-BSR}
f_\text{BSR}(m^2)=\frac {K}{\left[\mu^{2}(m^2)-M_X^{2}\right]^{2}+M_X^{2}\Gamma_X ^{2}}.
\end{equation}

In the Appendix it is shown that the choice $\epsilon=-1$ leads to a double peak at 
\begin{equation}
{m^*_{\pm}}^2=\frac{\Lambda^2}{2}\left[1\pm\left(1-4\frac{M_X^2}{\Lambda^2}\right)^{1/2} \right],
\end{equation} 
with widths
\begin{equation}
{\Gamma_\pm^*}^2=\frac{M_X^2\Gamma_X^2}{{m_\pm^*}^2\left(1-4{M_X^2}/{\Lambda^2}\right)}.
\end{equation}
This equation reveals a relationship between the widths of the two peaks,
\begin{equation}\label{eq:rel-peaks}
\frac{{\Gamma_+^*}^2}{{\Gamma_-^*}^2}=\frac{{m_-^*}^2}{{m_+^*}^2},
\end{equation}
which will be crucial in the distinction between the phenomenology of this double peak (\emph{twin peaks}) in a BSR scenario and the presence of two different resonances. 

In the next section we will consider this situation and apply it to the case of the $Z$-boson physics, which will allow us to put a lower bound on the scale $\Lambda$, and to the case of scattering in a very high energy hadron collider.

\section{Searches for BSR resonances in colliders}
\label{sec:limits}

\subsection{Bounds on $\Lambda$ using LEP data}
\label{sec:limits_1}

The main argument to take the $Z$ boson as our object of study is the precision achieved in the determination of its mass and decay width at the LEP collider~\cite{PhysRevD.98.030001}:
\begin{equation}
M_Z^{\text{\text{exp}}}=91,1876\pm 0.0021\,\mathrm{GeV},\quad\quad \Gamma_Z^{\text{exp}}=2.4952\pm 0.0023\,\mathrm{GeV},
\end{equation}
where the symbols of mass and width include the superscript ``exp'' to remark that these are the values that are obtained from a fit to a standard Breit-Wigner distribution, but that in the presence of BSR corrections would be different from the true values $M_Z$ and $\Gamma_Z$. $M_Z^2$ is the value of $\mu^2$, and $(M_Z^{\text{exp}})^2$, the value of $m^2$, at the peak of the Breit-Wigner distribution. Then one has, according to Eq.~\eqref{eq:MCL}:
\begin{equation}
M_Z^2=(M_Z^{\text{exp}})^2\left(1+\epsilon\frac{(M_Z^{\text{exp}})^2}{\Lambda^2} \right).
\end{equation}

The new value of $M_Z$ extracted from LEP data must be consistent with other determinations of $M_Z$, which can be used to put an upper bound to $\Lambda$. It is not trivial to calculate the maximum modification (${M}_Z=M_Z^{\text{exp}}+\delta M_Z$) of the $Z$ mass without spoiling the SM framework, but one can nevertheless put limits to this scale given some guess for this modification:\footnote{In the derivation of the bound on $\Lambda$ we are just comparing LEP data on the energy dependence of the cross section with the approximation around the first of the peaks in the generalized Breit-Wigner distribution. We have neglected the tiny effects of the tail of the second peak, which could be incorporated in a more systematic statitistical analysis. Uncertainties in the modifications due to the dynamics make such analysis not worthwhile at present.}
\begin{equation}\label{Lambda0}
\Lambda\geq \Lambda_0=M_Z^{\text{exp}}\left(\frac{M_Z^{\text{exp}}}{2\,\delta M_Z}\right)^{1/2} = 3,55 \,\left(\frac{30\,\text{MeV}}{\delta M_Z}\right)^{1/2} \,\text{TeV}.
\end{equation} 
Note that an energy scale of a few TeV, which is indeed within reach of future particle accelerators, is compatible with the bounds from LEP data. In the following subsection we will turn our attention to a future very high energy (VHE) hadron collider and will give a more detailed implementation of the idea for this case.

\subsection{Searches for BSR resonances in a VHE hadron collider}
\label{sec:limits_2}

LEP observations of the $Z$ boson do not show effects from the possible modification of SR considered in the previous model. A way to understand it in the context of the model is that the mass of the $Z$ boson is not close enough to the scale of new physics to show these effects. From the limits on the BSR scale [see Eq.~\eqref{Lambda0}], one can see that the LHC has not either sufficient energy to produce a particle with a mass which is large enough to allow for the identification of the new physics.

Given the lower bound on $\Lambda$ from LEP (a few TeVs), we can anticipate that the energy of a future electron-positron collider like ILC will not be enough to observe the two peaks of a new high-mass resonance from physics beyond the Standard Model. Let us then suppose some future VHE hadron (proton-proton, pp) collider which were able to observe the two peaks of a new resonance at $m^2={m_\pm^*}^2$ (we consider the interesting case $\epsilon=-1$), and let us assume that the resonance is produced through the annihilation of a quark and an antiquark of momenta $p$ and $\overline{p}$, respectively, and that decays to two fermions of momenta $q$ and $\overline{q}$, although there will be of course much more particles in the final state as a result of the hadron scattering. From Eq.~\eqref{BSR exp}, the differential cross section with respect to the variable $m^2=(q+\overline{q})^2$ would be
\begin{equation} \label{eq:difcs}
\frac{d\sigma}{dm^2}\approx \mathcal{F}_\pm(s,{m_\pm^*}^2) \frac{K_\pm}{(m^2-{m_\pm^*}^2)^2+{m_\pm^*}^2{\Gamma_\pm^*}^2}
\end{equation}
at each of the peaks, where the function $\mathcal{F}_\pm(s,{m_\pm^*}^2)$ can be determined through the parton model in the following way.

We will make use of the conventional Mandelstam variable $s$ of the proton-proton relativistic system, 
\begin{equation}
s=(P+\overline{P})^2=2E\overline{E}(1-\cos\theta)= 4E\overline{E},
\end{equation}
where $P$ and $\overline{P}$ are the four-momenta of the two protons in the initial state, and we have used $\theta=\pi$. We can write
\begin{equation}
P^\mu=\frac{\sqrt{s}}{2}(1,0,0,1),\quad\quad\quad\quad \overline{P}^\mu=\frac{\sqrt{s}}{2}(1,0,0,-1),
\end{equation}
for the four-momenta of the protons, and
 \begin{equation}
p^\mu=xP^\mu,\quad\quad\quad \overline{p}^\mu=\overline{x}\overline{P}^\mu,
\end{equation}
($0<x,\overline{x}<1$) for the four-momenta of the quark-antiquark pair that produce the resonance, which carry a certain fraction of the momentum of each one of the protons.

The squared of the invariant mass in SR of the quark-antiquark system is again, according to Eq.~\eqref{eq:s2},
\be
m^2=(p+\overline{p})^2=4E_p E_{\overline{p}}=4 x \overline{x} E \overline{E} = x \overline{x} s,
\label{eq:mxs}
\ee
where we have used the same symbol ($m^2$) at the initial and final states because it is a conserved quantity, that is, $(p+\overline{p})^2=(q+\overline{q})^2$ even though the MCL would be associated to a different conservation law, $(p\oplus\overline{p})^2=(q\oplus\overline{q})^2$, that is, the conservation of $\mu^2$. The reason is that the conservation of $\mu^2$ guarantees, through Eq.~\eqref{eq:MCL}, the conservation of $m^2$.

The relation $m^2=x\overline{x}s$ from Eq.~\eqref{eq:mxs} allows us to write the functions $\mathcal{F}_\pm(s,{m_\pm^*}^2)$ as
\begin{equation}\label{eq:F}
\mathcal{F}_\pm(s,{m_\pm^*}^2)=\int_0^1\int_0^1 dx\, d\overline{x} \,f_q(x,{m_\pm^*}^2) \, f_{\overline{q}}(\overline{x},{m_\pm^*}^2) \, \delta\left(x\overline{x}-\frac{{m_\pm^*}^2}{s}\right),
\end{equation}
where $f_q(x,{m_\pm^*}^2)$ is the parton distribution function (PDF), which is defined as the probability density to find a parton (quark) in a hadron (proton) with a fraction $x$ of its momentum when probing the hadron at an energy scale $m^2 \sim {m_\pm^*}^2$, where ${m_\pm^*}^2$ is given by Eq.~\eqref{epsilon-}.

The $K_\pm$ factors in Eq.~\eqref{eq:difcs} include the dependence of the couplings and every detail of the quark-antiquark annihilation. 
The previous expressions can be used to estimate the expected number of events for different pp colliders, different resonance mass values ($M_X$) and different values of the scale of new BSR physics ($\Lambda$). On the other hand, the observation of the double peak would allow to extract the true mass and width of the resonance through Eqs.~\eqref{changes} and~\eqref{eq:width}.

\section{Cross-section calculation in a quantum-field-theory approach beyond SR}
\label{sec:cross-section}

As commented above, the main objective of this work was to remark the possibility to explore scenarios beyond SR in accelerator physics, and to give some intuition on the expected signals from a simple model. In Sec.~\ref{sec:twin-peaks}, we considered a DSR model with a covariant MCL, and obtained the nontrivial result that a resonance may produce a pair of twin peaks in a cross-section distribution. This result, however, was based on the ansatz expressed by Eq.~\eqref{eq:BW-BSR}. 

In the present section we will try to make plausible this ansatz through an explicit cross-section calculation, taking the paradigmatic process $e^-(k) e^+(\overline{k})\rightarrow Z \rightarrow \mu^-(p) \mu^+(\overline{p})$ as an example, and with a MCL taken from the literature of DSR. To carry out the computation, we will use modified Feynman rules in such a way that they incorporate the modified kinematics through the substitution of the standard composition of momenta in the Mandelstam variables by the new composition given by the MCL. A few other ambiguities or adhoc prescriptions like this one will also be unavoidable, since a calculation from first principles would require a full dynamic QFT theory compatible with DSR kinematics, which is unknown at the moment. Nevertheless, this example will help to understand Eq.~\eqref{eq:BW-BSR} as a reasonable guess, as well as to illustrate the problem of \emph{channels}, which is generically present in a modification of the standard kinematics through a non-linear composition law.

\subsection{Phase-space momentum integrals}
\label{integral}

Let us first see the generalization of the two-particle phase-space integral in SR for the massless case:
\be\label{PS2}
PS_2 \,=\, \int \frac{d^4p}{(2\pi)^3}  \delta(p^2) \theta(p_0) \,\frac{d^4\overline{p}}{(2\pi)^3} \delta(\overline{p}^2) \theta(\overline{p}_0) \,(2\pi)^4 \delta^{(4)}[(k+\overline{k}) - (p+\overline{p})] \,. 
\ee
We will consider a BSR kinematics based on a noncommutative Lorentz covariant spacetime (Snyder algebra, Ref.~\cite{Battisti:2010sr}), for which a MCL for momenta appears, given by 
\begin{equation}\label{BSRMCL}
\left[l\oplus q\right]^{\mu}=l^{\mu}\sqrt{1+\frac{q^{2}}{\overline{\Lambda}^{2}}}+\frac{1}{\overline{\Lambda}^{2}\left(1+\sqrt{1+{l^{2}}/{\overline{\Lambda}^{2}}}\right)}l^{\mu}\left(l\cdot q\right)+q^{\mu}\approx l^{\mu}+q^{\mu}+\frac{1}{2\overline{\Lambda}^{2}}l^{\mu}\left(l \cdot q\right),
\end{equation}
where squared terms are neglected because we are considering relativistic particles. When one computes $(l\oplus q)^2$ from Eq.~\eqref{BSRMCL}, one gets the same expression obtained in Sec.~\ref{sec:twin-peaks} with $\epsilon=+1$.\footnote{The negative parameter $\epsilon=-1$ can be also considered in Eq.~\eqref{BSRMCL} if in the Snyder commutator of space-time coordinates we add both possibilities: $\left[x^\mu, x^\nu \right]=\pm J^{\mu\nu}/\overline{\Lambda}^2$. From now on, we will use Eq.~\eqref{BSRMCL}. Note also that Eq.~\eqref{eq:MCL} is recovered with $\Lambda^2=2\overline{\Lambda}^2$.} We have then a justification of the introduced model based on the simplest choice of a noncommutative spacetime.\footnote{Another example of DSR kinematics very much studied in the literature, known as $\kappa$-Minkowski spacetime (Ref.~\cite{Majid:1995qg}), has associated a noncovariant composition law, which would make the computation more difficult.} 

The generalization of the phase-space integral in the particular case of BSR where one has the dispersion relation of SR is defined by the MCL. For a non-symmetric MCL there are four different possible conservation laws and then  
four different ways (channels) in which the process $e^-(k) e^+(\overline{k}) \to Z \to \mu^-(p) \mu^+(\overline{p})$ can be produced. For each channel ($\alpha=1,2,3,4$), one has a generalized phase space integral
\be\label{PS2bar}
\overline{PS}_2^{(\alpha)} \,=\, \int \frac{d^4p}{(2\pi)^3}  \delta(p^2) \theta(p_0) \,\frac{d^4\overline{p}}{(2\pi)^3} \delta(\overline{p}^2) \theta(\overline{p}_0) \,(2\pi)^4 \delta^{(4)}_\alpha(k, \overline{k}; p, {\overline p}) ,
\ee
where
\begin{align}
\delta^{(4)}_1(k, \overline{k}; p, {\overline p}) &=\, \delta^{(4)}[(k\oplus\overline{k}) - (p\oplus\overline{p})], &
\delta^{(4)}_2(k, \overline{k}; p, {\overline p}) &=\, \delta^{(4)}[(\overline{k}\oplus k) - (p\oplus\overline{p})], \\
\delta^{(4)}_3(k, \overline{k}; p, {\overline p}) &=\, \delta^{(4)}[(k\oplus\overline{k}) - (\overline{p}\oplus p)], &
\delta^{(4)}_4(k, \overline{k}; p, {\overline p}) &=\, \delta^{(4)}[(\overline{k}\oplus k) - (\overline{p}\oplus p)].
\end{align}    
  
\subsection{Choice of the dynamical factor with a MCL}
\label{sec:dynamical}

Since the scattering takes place in a collider, we can assume that the particles in the initial state come from opposite directions, and we will make use of the ultra-relativistic limit (masses can be neglected): 
\be
k_\mu=\left(E_{0},\,\vec{k}\right),\, \overline{k}_\mu=\left(E_{0},\,-\vec{k}\right),\mbox{ with }E_{0}=|\vec{k}|.
\ee

The SR cross section at lowest order is given by a kinematic factor from the initial state, times the two-particle phase-space integral with an integrand which is a product of the squared modulus of the (unstable) $Z$-boson propagator and a dynamical factor $A$ from the coupling of the $Z$ boson to the particles in the initial and final states:
\be\label{sigma}
\sigma \,=\, \frac{1}{8 E_0^2} \, PS_2 \,\frac{1}{\left[(s - M_Z^2)^2 + \Gamma_Z^2 M_Z^2\right]} \,A.
\ee
The dynamical factor is given by~\cite{Atsue2015}:
\be\label{A}
A \,=\, \frac{e^4}{2 \sin^4\theta_W \cos^4\theta_W} \, \left(\left[C_V^2+C_A^2\right]^2 \left[\left(\frac{t}{2}\right)^2 + \left(\frac{u}{2}\right)^2\right] - 4 C_V^2 C_A^2 \left[\left(\frac{t}{2}\right)^2 - \left(\frac{u}{2}\right)^2\right]\right),
\ee
in terms of the corrections to the vector ($C_V$) and axial ($C_A$) weak charges, the Weinberg angle ($\theta_W$) and the Mandelstam variables 
\begin{align}
s&=\left(k+\overline{k}\right)^{2}=\left(p+\overline{p}\right)^{2}, \\
t&=\left(k-p\right)^{2}=\left(\overline{p}-\overline{k}\right)^{2}, \\
u&=\left(k-\overline{p}\right)^{2}=\left(p-\overline{k}\right)^{2}.
\end{align}
We do not have a generalization of RQFT compatible with the MCL and then we do not know what is the generalization of the SR cross section (\ref{sigma}). All we can do is a guess for such generalization compatible with Lorentz invariance which reduces in the limit $(E_0/\Lambda) \to 0$ to the SR cross section $\sigma$ in Eq.~(\ref{sigma}). The generalization of the two-particle phase-space integral in Eq.~(\ref{PS2bar}) leads to consider 
\be\label{sigmaalpha}
\overline{\sigma}_\alpha \,=\, \frac{1}{8 E_0^2} \, \overline{PS}_2^{(\alpha)} \,\frac{1}{\left[(\overline{s} - \overline{M}_Z^2)^2 + \overline{\Gamma}_Z^2 \overline{M}_Z^2\right]} \,\overline{A}_\alpha.
\ee
for each channel $\alpha$.\footnote{Note that the generalization of the squared total mass based on our choice for the MCL is the same for all the channels:
\be
(k\oplus\overline{k})^2 = (\overline{k}\oplus k)^2 = (p\oplus\overline{p})^2 = (\overline{p}\oplus p)^2 \doteq \overline{s}.
\ee}

In order to illustrate the uncertainty due to the lack of a dynamic framework, we consider two different guesses for the dynamical factor $\overline{A}_\alpha$: 
\begin{enumerate}
\item The simplest option is to assume that the dynamical factor written in terms of the invariants $t$, $u$ is independent of the MCL, $\overline{A}_\alpha=A$. But the MCL implies that $(k-p)\neq (\overline{p}-\overline{k})$, $(k-\overline{p})\neq (p-\overline{k})$; then one has to consider
\begin{align}
t &=  \frac{1}{2} \left[(k-p)^2 + (\overline{p} - \overline{k})^2\right] = - k\cdot p - \overline{k}\cdot \overline{p}, \\
u &=\, \frac{1}{2} \left[(k-\overline{p})^2 + (p-\overline{k})^2\right] = - k\cdot \overline{p} - \overline{k}\cdot p.
\end{align} 
\item Another possibility is to consider a generalization of $A$ based on the replacement of the Mandelstam variables $t$, $u$ by new invariants $\overline{t}$, $\overline{u}$. We cannot find a prescription to associate new invariants for each channel; we are then led to consider a channel independent ($\overline{A}_\alpha=\overline{A}$) dynamical factor which is obtained from $A$ by the replacement of the Mandelstam variables $t$, $u$ by\footnote{Note that in this example the squared of a composition of two momenta is symmetric even though the MCL is not symmetric.} 
\begin{align}
\overline{t} &= \frac{1}{2} \left[(k\oplus\hat{p})^2 + (\overline{p}\oplus \hat{\overline{k}})^2\right] = - k\cdot p - \overline{k}\cdot \overline{p} + \frac{(k\cdot p)^2}{2\overline{\Lambda}^2} + \frac{(\overline{k}\cdot \overline{p})^2}{2\overline{\Lambda}^2} ,\\
\overline{u} &= \frac{1}{2} \left[(k\oplus\hat{\overline{p}})^2 + (p\oplus \hat{\overline{k}})^2\right] = - k\cdot \overline{p} - \overline{k}\cdot p + \frac{(k\cdot \overline{p})^2}{2\overline{\Lambda}^2} + \frac{(\overline{k}\cdot p)^2}{2\overline{\Lambda}^2},
\end{align}
where we have used the notation \textit{antipode} $\hat{p}$ for the momentum whose composition with $p$ is zero (see Ref.~\cite{Majid:1995qg}). For the particular case of MCL given by Eq.~\eqref{BSRMCL}, and neglecting masses, $\hat{p}$ corresponds to a momentum whose components are the same as those of $p$, but with reversed sign.\footnote{Indeed,
\[
\begin{split}
\left[p\oplus \hat{p}\right]^{\mu}=\left[p\oplus -p\right]^{\mu}&= p^\mu \left\lbrace \sqrt{1+\frac{p^{2}}{\overline{\Lambda}^{2}}}-\frac{p^2}{\overline{\Lambda}^{2}\left(1+\sqrt{1+{p^{2}}/{\overline{\Lambda}^{2}}}\right)}-1 \right\rbrace  \\
&= p^\mu\left\lbrace \frac{\overline{\Lambda}^2\left( \sqrt{1+p^2/\overline{\Lambda}^2}+1\right)+\overline{\Lambda}^2 p^2/\overline{\Lambda}^2 -p^2}{\overline{\Lambda}^{2}\left(1+\sqrt{1+p^2/\overline{\Lambda}^2}\right)}-1 \right\rbrace = 0. 
\end{split}
\]}
\end{enumerate}

We do not know what is the channel that produces each final state. Then, in the comparison of the BSR model with the distribution of data as a function of $E_0^2$ we have to average over all channels and consider a cross section 
\be\label{sigmabar}
\overline{\sigma} \doteq \frac{1}{4}\sum_\alpha \overline{\sigma}_\alpha,
\ee 
with $\overline{\sigma}_\alpha$ in Eq.~(\ref{sigmaalpha}). In fact, one has two guesses for such cross section corresponding to the two choices discussed previously for the generalized dynamical factor $\overline{A}$
\begin{align}
\overline{\sigma}^{(1)} &= \frac{1}{32 E_0^2} \,\frac{1}{\left[(\overline{s} - \overline{M}_Z^2)^2 + \overline{\Gamma}_Z^2 \overline{M}_Z^2\right]} \,\sum_\alpha \overline{PS}_2^{(\alpha)} \,A(t,u),  \\
\overline{\sigma}^{(2)} &= \frac{1}{32 E_0^2} \,\frac{1}{\left[(\overline{s} - \overline{M}_Z^2)^2 + \overline{\Gamma}_Z^2 \overline{M}_Z^2\right]} \,\sum_\alpha \overline{PS}_2^{(\alpha)} \,A(\overline{t},\overline{u}).
\end{align}

We proceed to calculate the final expressions of the cross sections for these two cases in the following subsection.

\subsection{Cross sections with a MCL} 
\label{cross-sections}

To determine the cross section of the process $e^-(k) e^+(\overline{k}) \rightarrow Z \rightarrow \mu^-(p) \mu^+(\overline{p})$ with a MCL, one needs the two-particle phase space integral 
\be
\widehat{F}^{(\alpha)}(E_0) \doteq \overline{PS}_2^{(\alpha)} F(k, \overline{k}, p, \overline{p})
\ee
for different Lorentz invariant functions $F$ of the four momenta $k$, $\overline{k}$, $p$, $\overline{p}$. A first step is to use the Dirac delta function $\delta_\alpha^{(4)}(k, \overline{k}; p, \overline{p})$ corresponding to the conservation law for each channel to express $\overline{p}$ as a function $\overline{p}^{(\alpha)}(k, \overline{k}, p)$ of the remaining three momenta $k$, $\overline{k}$, $p$. Then, we have
\be
\widehat{F}^{(\alpha)}(E_0) = \frac{1}{(2\pi)^2} \int d^4p \,\delta(p^2) \theta(p_0) \,\delta\left(\overline{p}^{(\alpha)\,2}(k, \overline{k}, p)\right) \theta\left(\overline{p}_0^{(\alpha)}(k, \overline{k}, p)\right) F_\alpha(k, \overline{k}, p),
\ee
where
\be
F_\alpha(k, \overline{k}, p) = F(k, \overline{k}, p, \overline{p}^{(\alpha)}(k, \overline{k}, p)).
\ee
Next, we can integrate over $p_0$ and $|\vec{p}|$ with the remaining two Dirac delta functions
\be
\widehat{F}^{(\alpha)}(E_0) = \frac{1}{8\pi^2} \int d\Omega_{\hat{p}} \frac{E^{(\alpha)}(k, \overline{k}, \hat{p})}{|\frac{\partial\overline{p}^{(\alpha)\,2}}{\partial p_0}|_{p_0=E^{(\alpha)}(k, \overline{k}, \hat{p})}} \,F_\alpha(k, \overline{k}, p)|_{|\vec{p}|=p_0=E^{(\alpha)}(k, \overline{k}, \hat{p})}\,,
\ee
where $E^{(\alpha)}(k, \overline{k}, \hat{p})$ is the positive value of $p_0$ such that $\overline{p}^{(\alpha)\,2}=0$. Rotational invariance and the choice $\vec{\overline{k}}=-\vec{k}$ can be used to show that $E^{(\alpha)}$ is a function of the energy $E_0$ of the particles in the initial state and the angle $\theta$ between the directions of $\vec{k}$ and $\vec{p}$. Then we have
\be
\widehat{F}^{(\alpha)}(E_0) = \frac{1}{4\pi} \int d\cos\theta \frac{E^{(\alpha)}(E_0, \cos\theta)}{|\frac{\partial\overline{p}^{(\alpha)\,2}}{\partial p_0}|_{p_0=E^{(\alpha)}(E_0, \cos\theta)}} F_\alpha(k, \overline{k}, p)|_{|\vec{p}|=p_0=E^{(\alpha)}(E_0, \cos\theta)}\, .
\label{eq:Falpha}
\ee
We have to use now the explicit form of the conservation law in each channel to determine $E^{(\alpha)}(E_0, \cos\theta)$ and $|\frac{\partial\overline{p}^{(\alpha)\,2}}{\partial p_0}|_{p_0=E^{(\alpha)}(E_0, \cos\theta)}$.  

For the first channel, we have $k\oplus\overline{k}=p\oplus\overline{p}$, and then 
\be
k_\mu + \overline{k}_\mu + \frac{k\cdot \overline{k}}{2\overline{\Lambda}^2} k_\mu = 
p_\mu + \overline{p}_\mu + \frac{p\cdot \overline{p}}{2\overline{\Lambda}^2} p_\mu.
\ee
This implies that
\be
p\cdot \overline{p}^{(1)} = k\cdot p + \overline{k}\cdot p + \frac{(k\cdot \overline{k})(k\cdot p)}{2\overline{\Lambda}^2}
\ee
and, neglecting terms proportional to $(1/\overline{\Lambda}^4)$, one has
\be
\overline{p}^{(1)}_\mu = k_\mu + \overline{k}_\mu - p_\mu + \frac{k\cdot \overline{k}}{2\overline{\Lambda}^2} k_\mu - \frac{(k\cdot p+\overline{k}\cdot p)}{2\overline{\Lambda}^2} p_\mu,
\ee
and
\begin{align}
\overline{p}^{(1)\,2} &= 2 k\cdot \overline{k} - 2 k\cdot p - 2 \overline{k}\cdot p + \frac{(k\cdot \overline{k})^2}{\overline{\Lambda}^2} - \frac{(k\cdot \overline{k})(k\cdot p)}{\overline{\Lambda}^2} - \frac{(k\cdot p + \overline{k}\cdot p)^2}{\overline{\Lambda}^2}, \\
k\cdot \overline{p}^{(1)} &= k\cdot \overline{k} - k\cdot p - \frac{(k\cdot p+\overline{k}\cdot p)k\cdot p}{2\overline{\Lambda}^2}, \\
\overline{k}\cdot \overline{p}^{(1)} &= k\cdot \overline{k} - \overline{k}\cdot p + \frac{(k\cdot \overline{k})^2}{2\overline{\Lambda}^2} - \frac{(k\cdot p+\overline{k}\cdot p)\overline{k}\cdot p}{2\overline{\Lambda}^2} .
\end{align}
In the reference frame where $k_\mu = E_0 (1, \hat{k})$, $\overline{k}_\mu = E_0 (1, -\hat{k})$, one has 
\begin{align}
\overline{p}^{(1)\,2} &= 4 E_0^2 - 4 E_0 p_0 + \frac{4E_0^4}{\overline{\Lambda}^2} - \frac{2E_0^3}{\overline{\Lambda}^2} p_0 (1-\cos\theta) - \frac{4E_0^2}{\overline{\Lambda}^2} p_0^2, \\
k\cdot \overline{p}^{(1)} &= 2 E_0^2 - E_0 p_0 (1-\cos\theta) - \frac{E_0^3}{\overline{\Lambda}^2} p_0 (1-\cos\theta), \\
\overline{k}\cdot \overline{p}^{(1)} &= 2E_0^2 - E_0 p_0 (1+\cos\theta) + \frac{2E_0^4}{\overline{\Lambda}^2} - \frac{E_0^2}{\overline{\Lambda}^2} p_0^2 (1+\cos\theta).
\end{align}
From the expression of $\overline{p}^{(1)\,2}$, one finds 
\be
\frac{\partial \overline{p}^{(1)\,2}}{\partial p_0} = - 4E_0 - \frac{2E_0^3}{\overline{\Lambda}^2} (1-\cos\theta) - \frac{8E_0^2}{\overline{\Lambda}^2} p_0,\quad\quad\quad E^{(1)}=E_0\left(1 - \frac{E_0^2}{2\overline{\Lambda}^2} (1-\cos\theta)\right).
\ee 
A similar analysis can be made for the other three channels.
 
To calculate the cross section, we need to consider four invariant functions
\be
\label{eq:Finv}
F_{\pm}=t^2 \pm u^2=(k\cdot p + \overline{k}\cdot \overline{p})^2 \pm (k\cdot \overline{p} + \overline{k}\cdot p)^2 ,
\ee
\be
\label{eq:F-inv}
\begin{split}
\overline{F}_{\pm}=\overline{t}^2 \pm \overline{u}^2=&\left[(k\cdot p + \overline{k}\cdot \overline{p})^2 - (k\cdot p + \overline{k}\cdot \overline{p})\left[(k\cdot p)^2+(\overline{k}\cdot \overline{p})^2\right]/\overline{\Lambda}^2\right] \\
&\pm \left[(k\cdot \overline{p} + \overline{k}\cdot p)^2 - (k\cdot \overline{p} + \overline{k}\cdot p)\left[(k\cdot \overline{p})^2+(\overline{k}\cdot p)^2\right]/\overline{\Lambda}^2\right] ,
\end{split}
\ee
and the corresponding phase-space integrals $\widehat{F}_{\pm}^{(\alpha)}(E_0)$, $\widehat{\overline{F}}_{\pm}^{(\alpha)}(E_0)$. The cross sections with a MCL for the two guesses for the dynamical factor $A$ described in Sec.~\ref{sec:dynamical} are:
\begin{align}
\overline{\sigma}^{(1)} &= \frac{e^4}{256 \sin^4\theta_W\cos^4\theta_W E_0^2} \,\frac{1}{\left[(\overline{s} - \overline{M}_Z^2)^2 + \overline{\Gamma}_Z^2 \overline{M}_Z^2\right]} \, \left[(C_V^2 + C_A^2)^2 \sum_\alpha \widehat{F}_+^{(\alpha)}(E_0) - 4 C_V^2 C_A^2 \sum_\alpha \widehat{F}_-^{(\alpha)}(E_0)\right], \\
\overline{\sigma}^{(2)} &= \frac{e^4}{256 \sin^4\theta_W\cos^4\theta_W E_0^2} \,\frac{1}{\left[(\overline{s} - \overline{M}_Z^2)^2 + \overline{\Gamma}_Z^2 \overline{M}_Z^2\right]} \, \left[(C_V^2 + C_A^2)^2 \sum_\alpha \widehat{\overline{F}}_+^{(\alpha)}(E_0) - 4 C_V^2 C_A^2 \sum_\alpha \widehat{\overline{F}}_-^{(\alpha)}(E_0)\right] .
\end{align}

Upon substitution of  the results for the phase-space integrals $\widehat{F}^{(\alpha)}_\pm(E_0)$, $\widehat{\overline{F}}^{(\alpha)}_\pm(E_0)$, obtained by applying Eq.~\eqref{eq:Falpha} to the four invariants in Eqs.~\eqref{eq:Finv}-\eqref{eq:F-inv}, we get the final results for these cross sections:
\begin{align}
  \overline{\sigma}^{(1)} &= \frac{e^4}{48 \pi \sin^4\theta_W \cos^4\theta_W} \,\frac{E_0^2}{\left[\left(4E_0^2(1+E_0^2/\overline{\Lambda}^2) - \overline{M}_Z^2\right)^2 + \overline{\Gamma}_Z^2 \overline{M}_Z^2\right]} \, \left((C_V^2+C_A^2)^2 \left[1 -\frac{E_0^2}{2\Lambda ^2}\right] \right),\\
  \overline{\sigma}^{(2)} &=  \frac{e^4}{48 \pi \sin^4\theta_W \cos^4\theta_W} \,\frac{E_0^2}{\left[\left(4E_0^2(1+E_0^2/\overline{\Lambda}^2) - \overline{M}_Z^2\right)^2 + \overline{\Gamma}_Z^2 \overline{M}_Z^2\right]} \,\left((C_V^2+C_A^2)^2 \left[1 -\frac{2 E_0^2}{\Lambda ^2}\right] \right).
\end{align}  

\subsection{Constraints to $\overline{\Lambda}$}

Let us now see the restrictions imposed by the cross section, taking into account the PDG data~\cite{PhysRevD.98.030001}. We require that there is a value of $\overline{M}_Z$ and $\overline{\Gamma}_Z$, in an interval\footnote{One can see that bigger values for $\delta\overline{M}_Z$ and $\delta\overline{\Gamma}_Z$ do not vary significantly the constraint for $\overline{\Lambda}$.} $\pm 30\,\text{MeV}$ around their central values given by the PDG, making the cross section $\overline{\sigma}$ compatible with data. Given the success of the Standard Model, we take the SR cross section, with the PDG values of $M_Z$ and $\Gamma_Z$ at one or two standard deviations, as a good approximation to the data. 

We show the results in Table~\ref{table:two}, where we denote by $\overline{\sigma}^{(j)}_{i}$ the cross section taking $i$ standard deviations in the data.\footnote{Notice that the results in Table~\ref{table:two} are independent of the choice of sign in the MCL.}
\begin{table}
\begin{center}
\begin{tabular}{|c|c|c|c|c|c|c|c|c|}
\hline 
Constraints & $\overline{\sigma}^{(1)}_{1}$ & $\overline{\sigma}^{(1)}_{2}$ & $\overline{\sigma}^{(2)}_{1}$ & $\overline{\sigma}^{(2)}_{2}$  \tabularnewline
$\overline{\Lambda}${[}TeV{]} & 2.2 &1.8 & 2.5 & 1.8 \tabularnewline
\hline 
\end{tabular}
\par\end{center}
\caption{Bounds on the scale of new physics $\overline{\Lambda}$ from LEP data of the $Z$ boson.}
\label{table:two}
\end{table}

From the presented numerical values in the previous table we conclude that, in spite of the fact that the detailed calculations carried out in this last section allows us to understand better the new physics beyond SR in the framework of QFT, a good estimate is to take the simple approximation used in Sec.~\ref{sec:twin-peaks} [Eq.~\eqref{eq:MCL}], since the constraints imposed by that composition law barely differ from the ones given by the whole cross section.

\section{Summary and conclusions}
\label{sec:conclusions}

In this work, we have presented the attractive idea that footprints of a modification of SR driven by a low enough energy scale could be observed in accelerator physics experiments. As an example, a very simple model of MCL appearing in relativistic extensions of SR may show, depending on the sign of the correction to SR, a peculiar (and easily identifiable) signal (a pair of ``twin peaks'') which would show up in the study of new high-energy resonances.

The specific model studied in this work (Sec.~\ref{sec:twin-peaks}) takes as a departure point the Breit-Wigner distribution, which is maintained in the relativistic generalization of the kinematics of SR. The deformation of SR is introduced at the level of a modified composition law of momenta, which gives a new relation between the momentum of the resonance $q$ and the momenta $(k,\overline{k})$ of the particles that produce the resonance, $q=(k\oplus \overline{k})$, or with the momenta $p$, $\overline{p}$ of the particles in which the resonance decays, $q=p\oplus \overline{p}$.

For a composition law such that $(l\oplus q)^2=(l+q)^2[1-(l+q)^2/(2\overline{\Lambda}^2)]$, the Breit-Wigner distribution presents, as a function of $(l+q)^2$, two poles whose positions and widths are determined by the mass $M_X$ and width $\Gamma_X$ of the resonance, and the scale $\overline{\Lambda}$ of new physics. There is then a relationship between the position of the poles $(m_+^2,m_-^2)$ and the widths $(\Gamma_+,\Gamma_-)$. It is this relation, $m_ +\Gamma_+=m_-\Gamma_-$ [Eq.~\eqref{eq:rel-peaks}], what defines the unexpected (in SR) new kinematic effect (``twin peaks''). 

In the case of production of a resonance in two-particle scattering, the cross section will present the double peak as a function of $m^2=(k+\overline{k})^2$. In the case of observation a resonance (which is produced together with a number of other particles), through its decay to two particles of momenta $p$ and $\overline{p}$, the double peak will appear in the differential cross section, expressed as a function of $m^2=(p+\overline{p})^2$. The first case is relevant in the study of limits to the scale $\Lambda$, that can be extracted from LEP data on the $Z$ boson (Sec.~\ref{sec:limits_1}). The second case would be relevant in the search for effects from a deformed relativistic kinematics in a future hadron (pp) collider at $100$\,TeV (Sec.~\ref{sec:limits_2}).

Sections~\ref{sec:twin-peaks} and~\ref{sec:limits} contain the main results of this paper. In Sec.~\ref{sec:cross-section} we have also offered a specific cross-section calculation in the framework of QFT. The computation is however not free of some adhoc prescriptions, which is something unavoidable in the absence of a full dynamic QFT approach which should be consistent with the deformed kinematics. In the present work, we have assumed that the standard description of the production of resonances in relativistic QFT, given by the relativisic Breit-Wigner distribution, can be extended to a deformation compatible with relativistic invariance in which all the effect of the deformation is contained in the deformed expression of the energy-momentum of the resonance in terms of the momenta of the particles producing the resonance, or of the momenta of the particles produced in the decay of the resonance. The absence of a well-defined deformation of relativistic quantum field theory does not allow one to give a proof of the validity of such assumption, for which ideas of integrability~\cite{Mironov:2017gja} might offer a guiding principle. Such an extension of relativistic quantum field theory should be the objective of future development in the domain of DSR theories.

\authorcontributions{All authors contributed equally to the present work.}

\acknowledgments{This work is supported by the Spanish MINECO FPA2015-65745-P (MINECO/FEDER) and Spanish DGIID-DGA Grant No. 2015-E24/2.}

\conflictsofinterest{The authors declare no conflict of interest.} 

\abbreviations{The following abbreviations are used in this manuscript:\\

\noindent 
\begin{tabular}{@{}ll}
SR & Special Relativity \\
QFT & Quantum Field Theory \\
GR & General Relativity \\ 
LIV & Lorentz Invariant Violation\\
DSR & Deformed Special Relativity\\
LHC & Large Hadron Collider \\
LEP & Lepton Electron-Positron collider \\
MCL & Modified Composition Law \\
BSR & Beyond Special Relativity \\
VHE & Very High Energy \\
pp & proton-proton \\
PDG & Particle Data Group
\end{tabular}}

\appendixtitles{yes} 
\appendixsections{one} 
\appendix

\section{BSR extension of the Breit-Wigner distribution} 
\label{appendix:B-W}

From Eq.~\eqref{eq:BW-BSR},
\be
\label{eqapp:BW-BSR}
f_\text{BSR}(m^2)=\frac {K}{\left[\mu^{2}(m^2)-M_X^{2}\right]^{2}+M_X^{2}\Gamma_X ^{2}},
\end{equation}
it is convenient to introduce the dimensionless variables
\begin{equation}\label{dimensionless}
\tau:=\frac{m^2}{M_X^2}\quad \quad\quad \gamma:=\frac{\Gamma_X^2}{M_X^2}\quad\quad \quad \lambda:=\frac{\Lambda^2}{M_X^2},
\end{equation}
so that, once they are replaced in Eq.~\eqref{eqapp:BW-BSR}, one gets the expression
\begin{equation}
\label{simplification}
f_\text{BSR}(m^2)=\frac{K}{M_X^4F(\tau)},
\end{equation}
where
\begin{equation}
F(\tau):=\left[\tau\left(1+\epsilon\frac{\tau}{\lambda}\right)-1\right]^2+\gamma
\end{equation}
for the simplest choice of a BSR modified kinematics Eq.~\eqref{eq:MCL}.

In order to have a resonance, we need $\gamma\ll 1$, so the condition of a peak is 
\begin{equation}
\tau\left(1+\epsilon\frac{\tau}{\lambda}\right)-1=0,
\end{equation}
with solutions
\begin{equation}
\tau^*=-\frac{\lambda}{2\epsilon}\left[1\pm\left(1+4\frac{\epsilon}{\lambda}\right)^{1/2}\right].
\end{equation}

The distribution for values of $m^2$ near to the peaks can be analyzed by making a Taylor expansion at $\tau^*$. Evaluating the derivatives of $F(\tau)$ up to second order
\begin{equation}
\left.\frac{dF}{d\tau}\right\vert_{\tau=\tau^*}=2\left[\tau^*\left(1+\epsilon\frac{\tau^*}{\lambda}\right)-1 \right]\left(1+2\frac{\epsilon}{\lambda}\tau^*\right)=0,
\end{equation}
\begin{equation}
\left.\frac{d^2F}{d\tau^2}\right\vert_{\tau=\tau^*}=2\left(1+2\frac{\epsilon}{\lambda}\tau^* \right)^2=2\left(1+4\frac{\epsilon}{\lambda}\right),
\end{equation}
one obtains
\begin{equation}
F(\tau)\approx \gamma+\left(1+4\frac{\epsilon}{\lambda}\right)(\tau-\tau^*)^2,
\end{equation}
that substituted in Eq.~\eqref{simplification}, and using Eq.~\eqref{dimensionless}, leads to
\begin{equation}\label{BSR exp}
f_\text{BSR}(m^2)\approx \frac{1}{M^4_X\left(1+4\epsilon M_X^2/\Lambda^2 \right)}\cdot \frac{K}{{(m^2-{m^*}^2)}^2+M_X^2\Gamma_X^2 \left(1+4\epsilon M_X^2/\Lambda^2 \right)^{-1}}\,,
\end{equation}
where
\begin{equation}\label{m_*}
{m^*}^2:=M_X^2\tau^*=\frac{\Lambda^2}{2\epsilon}\left[-1\pm\left(1+4\epsilon\frac{M_X^2}{\Lambda^2}\right)^{1/2} \right].
\end{equation}

We see that the maximum value of the distribution~\eqref{BSR exp} is reached at $m^2={m^*}^2$, and one has then to consider separately if the parameter $\epsilon$ of the correction in Eq.~\eqref{eq:MCL} is positive or negative.

The choice $\epsilon=+1$ leads to a unique solution for the pole 
\begin{equation}\label{epsilon+}
{m^*}^2=\frac{\Lambda^2}{2}\left[\left(1+4\frac{M_X^2}{\Lambda^2}\right)^{1/2}-1 \right].
\end{equation}
The form of the distribution~\eqref{BSR exp} is the same as in SR, but the position of the peak $(m^2={m^*}^2)$ does not give us the squared mass of the resonance. It is easy to check from Eq.~\eqref{epsilon+} that when the resonance mass is much smaller than the scale of new physics, $M_X\ll \Lambda$,
\begin{equation}
{m^*}^2\approx\frac{\Lambda^2}{2}\left[1+2\frac{M_X^2}{\Lambda^2}-1 \right]=M_X^2,
\end{equation}
and we recover the peak of the SR distribution. 

From Eq.~\eqref{BSR exp}, the width of the peak will be [compare with the Breit-Wigner distribution~\eqref{eq:BW}]:
\begin{equation}
{\Gamma^*}^2=\frac{M_X^2\Gamma_X^2}{{m^*}^2\left(1+4{M_X^2}/{\Lambda^2}\right)}=\Gamma_X^2\frac{2{M_X^2}/{\Lambda^2}}{\left(1+4{M_X^2}/{\Lambda^2}\right)\left[\left(1+4{M_X^2}/{\Lambda^2}\right)^{1/2}-1 \right]},
\label{eq:width+}
\end{equation}
which also approaches the decay width of the resonance when $M_X^2\ll \Lambda^2$.

The $\epsilon=-1$ case turns out to be much more interesting. Assuming that
\begin{equation}
1-4\frac{M_X^2}{\Lambda^2}>0,
\end{equation}
that is, that the condition $M_X<\Lambda/2$ is satisfied, Eq.~\eqref{m_*} gives us two solutions for which ${m^*}^2>0$,
\begin{equation}\label{epsilon-}
{m^*_{\pm}}^2=\frac{\Lambda^2}{2}\left[1\pm\left(1-4\frac{M_X^2}{\Lambda^2}\right)^{1/2} \right].
\end{equation} 
This means that we have two peaks (at $m^2={m^*_{\pm}}^2$) in the squared mass distribution instead of one. From the position of these two peaks, we can get $\Lambda^2$ and $M_X^2$ using Eq.~\eqref{epsilon-}:
\begin{equation}\label{changes}
\Lambda^2=({m^*_+}^2+{m^*_-}^2)\,, \quad \quad\quad M_X^2=\frac{{m^*_+}^2{m^*_-}^2}{({m^*_+}^2+{m^*_-}^2)}\,.
\end{equation}

As in Eq.~\eqref{eq:width+}, the widths of the two peaks are
\begin{equation}\label{gamma1}
{\Gamma_\pm^*}^2=\frac{M_X^2\Gamma_X^2}{{m_\pm^*}^2\left(1-4{M_X^2}/{\Lambda^2}\right)}.
\end{equation}

Using the expressions of $M_X^2$ and $\Lambda^2$ in Eq.~\eqref{changes}, we get the combination
\be
1-4\frac{M_X^2}{\Lambda^2}=\frac{({m_+^*}^2-{m_-^*}^2)^2}{({m_+^*}^2+{m_-^*}^2)^2}.
\ee
Substitution in Eq.~\eqref{gamma1} gives then
\begin{equation}\label{gamma}
{\Gamma_\pm^*}^2=\Gamma_X^2\frac{({m_+^*}^2+{m_-^*}^2){m_\mp^*}^2}{({m_+^*}^2-{m_-^*}^2)^2}.
\end{equation}
Reversing Eq.~\eqref{gamma}, one can compute the decay width of the resonance $X$:
\begin{equation} \label{eq:width}
\Gamma_X^2={\Gamma_+^*}^2\frac{({m_+^*}^2-{m_-^*}^2)^2}{{m_-^*}^2({m_+^*}^2+{m_-^*}^2)}={\Gamma_-^*}^2\frac{({m_+^*}^2-{m_-^*}^2)^2}{{m_+^*}^2({m_+^*}^2+{m_-^*}^2)}=({\Gamma_+^*}^2+{\Gamma_-^*}^2)\left[\frac{{m_+^*}^2-{m_-^*}^2}{{m_+^*}^2+{m_-^*}^2}\right]^2.
\end{equation}

In the limit $M_X^2\ll\Lambda^2$, the expressions for the poles [Eq.~\eqref{epsilon-}] and the widths [Eq.~\eqref{gamma1}] become
\begin{equation}
{m^*_{\pm}}^2\approx\frac{\Lambda^2}{2}\left[1\pm \left(1-2\frac{M_X^2}{\Lambda^2}\right)\right],
\end{equation}
\begin{equation}
{\Gamma_\pm^*}^2\approx\Gamma_X^2\frac{2{M_X^2}/{\Lambda^2}}{\left[1\pm\left(1-2{M_X^2}/{\Lambda^2}\right) \right]},
\end{equation}
so that in this limit
\begin{equation}
\begin{array}{ll}
{m_+^*}^2\approx \Lambda^2 \,,& {\Gamma_+^*}^2\approx \Gamma_X^2 {M_X^2}/{\Lambda^2}\,, \\
{m_-^*}^2 \approx M_X^2 \,,& {\Gamma_-^*}^2\approx \Gamma_X^2\,,
\end{array}
\end{equation}
and we see that one of the peaks ($-$) reproduces the result of SR, while the other one ($+$) is displaced by a factor $\Lambda/M_X$, and its width reduced by a factor $M_X/\Lambda$, with respect to the SR peak. 

It is interesting to note that for $M_X>\Lambda/2$ we do not have any peak, since the term in the square root in Eq.~\eqref{epsilon-} gets negative. This would be the case of an ``invisible'' resonance. If the limit $M_X\rightarrow \Lambda/2$ is taken in the previous expressions, the two poles coincide, but their width tend to infinite. 

In the previous discussion we have not included the dependence on $m^2$ of the $K$ factor that takes into account the mechanism of production of the resonance and the decay width in the two particles which are observed. We are assuming that the analysis is valid in a small enough region around the peaks, where $K\approx K({m^*}^2)$, and the variation of $K$ with respect to $m^2$ can be neglected.


\begin{thebibliography}{-------}
\providecommand{\natexlab}[1]{#1}

\bibitem[Feynman(1996)]{Feynman:1996kb}
Feynman, R.P.
\newblock {\em {Feynman lectures on gravitation}};  1996.

\bibitem[Birrell and Davies(1984)]{Birrell:1982ix}
Birrell, N.D.; Davies, P.C.W.
\newblock {\em {Quantum Fields in Curved Space}}; Cambridge Monographs on
  Mathematical Physics, Cambridge Univ. Press: Cambridge, UK,  1984.

\bibitem[Bogoliubov and Shirkov(1982)]{Bogoliubov:1982book}
Bogoliubov, N.; Shirkov, D.
\newblock {\em Quantum Fields}; Addison-Wesley,  1982.

\bibitem[Mukhi(2011)]{Mukhi:2011zz}
Mukhi, S.
\newblock {String theory: a perspective over the last 25 years}.
\newblock {\em Class. Quant. Grav.} {\bf 2011}, {\em 28},~153001,
  \href{http://xxx.lanl.gov/abs/1110.2569}{{\normalfont
  [arXiv:physics.pop-ph/1110.2569]}}.

\bibitem[Aharony(2000)]{Aharony:1999ks}
Aharony, O.
\newblock {A Brief review of 'little string theories'}.
\newblock {\em Class. Quant. Grav.} {\bf 2000}, {\em 17},~929--938,
  \href{http://xxx.lanl.gov/abs/hep-th/9911147}{{\normalfont
  [arXiv:hep-th/hep-th/9911147]}}.

\bibitem[Dienes(1997)]{Dienes:1996du}
Dienes, K.R.
\newblock {String theory and the path to unification: A Review of recent
  developments}.
\newblock {\em Phys. Rept.} {\bf 1997}, {\em 287},~447--525,
  \href{http://xxx.lanl.gov/abs/hep-th/9602045}{{\normalfont
  [arXiv:hep-th/hep-th/9602045]}}.

\bibitem[Sahlmann(2010)]{Sahlmann:2010zf}
Sahlmann, H.
\newblock {Loop Quantum Gravity - A Short Review}.
\newblock  {Proceedings, Foundations of Space and Time: Reflections on Quantum
  Gravity: Cape Town, South Africa},  2010, pp. 185--210,
  \href{http://xxx.lanl.gov/abs/1001.4188}{{\normalfont
  [arXiv:gr-qc/1001.4188]}}.

\bibitem[Dupuis \em{et~al.}(2012)Dupuis, Ryan, and Speziale]{Dupuis:2012yw}
Dupuis, M.; Ryan, J.P.; Speziale, S.
\newblock {Discrete gravity models and Loop Quantum Gravity: a short review}.
\newblock {\em SIGMA} {\bf 2012}, {\em 8},~052,
  \href{http://xxx.lanl.gov/abs/1204.5394}{{\normalfont
  [arXiv:gr-qc/1204.5394]}}.

\bibitem[Van~Nieuwenhuizen(1981)]{VanNieuwenhuizen:1981ae}
Van~Nieuwenhuizen, P.
\newblock {Supergravity}.
\newblock {\em Phys. Rept.} {\bf 1981}, {\em 68},~189--398.

\bibitem[Taylor(1984)]{Taylor:1983su}
Taylor, J.G.
\newblock {A Review of Supersymmetry and Supergravity}.
\newblock {\em Prog. Part. Nucl. Phys.} {\bf 1984}, {\em 12},~1--101.

\bibitem[Wallden(2013)]{Wallden:2013kka}
Wallden, P.
\newblock {Causal Sets Dynamics: Review \& Outlook}.
\newblock {\em J. Phys. Conf. Ser.} {\bf 2013}, {\em 453},~012023.

\bibitem[Wallden(2010)]{Wallden:2010sh}
Wallden, P.
\newblock {Causal Sets: Quantum Gravity from a Fundamentally Discrete
  Spacetime}.
\newblock {\em J. Phys. Conf. Ser.} {\bf 2010}, {\em 222},~012053,
  \href{http://xxx.lanl.gov/abs/1001.4041}{{\normalfont
  [arXiv:gr-qc/1001.4041]}}.

\bibitem[Henson(2006)]{Henson:2006kf}
Henson, J.
\newblock {The Causal set approach to quantum gravity} {\bf 2006}.
\newblock pp. 393--413,
  \href{http://xxx.lanl.gov/abs/gr-qc/0601121}{{\normalfont
  [arXiv:gr-qc/gr-qc/0601121]}}.

\bibitem[Gross and Mende(1988)]{Gross:1987ar}
Gross, D.J.; Mende, P.F.
\newblock {String Theory Beyond the Planck Scale}.
\newblock {\em Nucl. Phys.} {\bf 1988}, {\em B303},~407--454.

\bibitem[Amati \em{et~al.}(1989)Amati, Ciafaloni, and Veneziano]{Amati:1988tn}
Amati, D.; Ciafaloni, M.; Veneziano, G.
\newblock {Can Space-Time Be Probed Below the String Size?}
\newblock {\em Phys. Lett.} {\bf 1989}, {\em B216},~41--47.

\bibitem[Garay(1995)]{Garay1995}
Garay, L.J.
\newblock {Quantum gravity and minimum length}.
\newblock {\em Int. J. Mod. Phys.} {\bf 1995}, {\em A10},~145--166,
  \href{http://xxx.lanl.gov/abs/gr-qc/9403008}{{\normalfont
  [arXiv:gr-qc/gr-qc/9403008]}}.

\bibitem[Wheeler(1955)]{Wheeler:1955zz}
Wheeler, J.A.
\newblock {Geons}.
\newblock {\em Phys. Rev.} {\bf 1955}, {\em 97},~511--536.

\bibitem[Ng(2011)]{Ng:2011rn}
Ng, Y.J.
\newblock {Various Facets of Spacetime Foam}.
\newblock  {Time and Matter: Proceedings, 3rd International Conference,
  TAM2010, Budva, Montenegro, 4-8 October, 2010},  2011, pp. 103--122,
  \href{http://xxx.lanl.gov/abs/1102.4109}{{\normalfont
  [arXiv:gr-qc/1102.4109]}}.

\bibitem[Feynman \em{et~al.}(1963)Feynman, Leighton, and
  Sands]{Feynman:1963uxa}
Feynman, R.P.; Leighton, R.B.; Sands, M., The Feynman Lectures on Physics;
  1963; chapter 38 (vol. 1).

\bibitem[Kostelecky and Russell(2011)]{Kostelecky:2008ts}
Kostelecky, V.A.; Russell, N.
\newblock {Data Tables for Lorentz and CPT Violation}.
\newblock {\em Rev. Mod. Phys.} {\bf 2011}, {\em 83},~11--31,
  \href{http://xxx.lanl.gov/abs/0801.0287}{{\normalfont
  [arXiv:hep-ph/0801.0287]}}.

\bibitem[Amelino-Camelia(2013)]{AmelinoCamelia:2008qg}
Amelino-Camelia, G.
\newblock {Quantum-Spacetime Phenomenology}.
\newblock {\em Living Rev.Rel.} {\bf 2013}, {\em 16},~5,
  \href{http://xxx.lanl.gov/abs/0806.0339}{{\normalfont
  [arXiv:gr-qc/0806.0339]}}.

\bibitem[Mattingly(2005)]{Mattingly:2005re}
Mattingly, D.
\newblock {Modern tests of Lorentz invariance}.
\newblock {\em Living Rev.Rel.} {\bf 2005}, {\em 8},~5,
  \href{http://xxx.lanl.gov/abs/gr-qc/0502097}{{\normalfont
  [arXiv:gr-qc/gr-qc/0502097]}}.

\bibitem[Liberati(2013)]{Liberati2013}
Liberati, S.
\newblock {Tests of Lorentz invariance: a 2013 update}.
\newblock {\em Class.Quant.Grav.} {\bf 2013}, {\em 30},~133001,
  \href{http://xxx.lanl.gov/abs/1304.5795}{{\normalfont
  [arXiv:gr-qc/1304.5795]}}.

\bibitem[Carmona \em{et~al.}(2018)Carmona, Cort\'es, and
  Relancio]{Carmona:2018xwm}
Carmona, J.M.; Cort\'es, J.L.; Relancio, J.J.
\newblock {Observers and their notion of spacetime beyond special relativity}.
\newblock {\em Symmetry} {\bf 2018}, {\em 10},~231,
  \href{http://xxx.lanl.gov/abs/1806.01725}{{\normalfont
  [arXiv:hep-th/1806.01725]}}.

\bibitem[Arkani-Hamed \em{et~al.}(1998)Arkani-Hamed, Dimopoulos, and
  Dvali]{ArkaniHamed:1998rs}
Arkani-Hamed, N.; Dimopoulos, S.; Dvali, G.R.
\newblock {The Hierarchy problem and new dimensions at a millimeter}.
\newblock {\em Phys. Lett.} {\bf 1998}, {\em B429},~263--272,
  \href{http://xxx.lanl.gov/abs/hep-ph/9803315}{{\normalfont
  [arXiv:hep-ph/hep-ph/9803315]}}.

\bibitem[Antoniadis \em{et~al.}(1998)Antoniadis, Arkani-Hamed, Dimopoulos, and
  Dvali]{Antoniadis:1998ig}
Antoniadis, I.; Arkani-Hamed, N.; Dimopoulos, S.; Dvali, G.R.
\newblock {New dimensions at a millimeter to a Fermi and superstrings at a
  TeV}.
\newblock {\em Phys. Lett.} {\bf 1998}, {\em B436},~257--263,
  \href{http://xxx.lanl.gov/abs/hep-ph/9804398}{{\normalfont
  [arXiv:hep-ph/hep-ph/9804398]}}.

\bibitem[Sundrum(1999)]{Sundrum:1998ns}
Sundrum, R.
\newblock {Compactification for a three-brane universe}.
\newblock {\em Phys. Rev.} {\bf 1999}, {\em D59},~085010,
  \href{http://xxx.lanl.gov/abs/hep-ph/9807348}{{\normalfont
  [arXiv:hep-ph/hep-ph/9807348]}}.

\bibitem[Arkani-Hamed \em{et~al.}(2001)Arkani-Hamed, Dimopoulos, and
  March-Russell]{ArkaniHamed:1998kx}
Arkani-Hamed, N.; Dimopoulos, S.; March-Russell, J.
\newblock {Stabilization of submillimeter dimensions: The New guise of the
  hierarchy problem}.
\newblock {\em Phys. Rev.} {\bf 2001}, {\em D63},~064020,
  \href{http://xxx.lanl.gov/abs/hep-th/9809124}{{\normalfont
  [arXiv:hep-th/hep-th/9809124]}}.

\bibitem[Tanabashi \em{et~al.}(2018)Tanabashi, Hagiwara, Hikasa, Nakamura,
  Sumino, Takahashi, Tanaka, Agashe, Aielli, Amsler, Antonelli, Asner, Baer,
  Banerjee, Barnett, Basaglia, Bauer, Beatty, Belousov, Beringer, Bethke,
  Bettini, Bichsel, Biebel, Black, Blucher, Buchmuller, Burkert, Bychkov, Cahn,
  Carena, Ceccucci, Cerri, Chakraborty, Chen, Chivukula, Cowan, Dahl,
  D'Ambrosio, Damour, de~Florian, de~Gouv\^ea, DeGrand, de~Jong, Dissertori,
  Dobrescu, D'Onofrio, Doser, Drees, Dreiner, Dwyer, Eerola, Eidelman, Ellis,
  Erler, Ezhela, Fetscher, Fields, Firestone, Foster, Freitas, Gallagher,
  Garren, Gerber, Gerbier, Gershon, Gershtein, Gherghetta, Godizov, Goodman,
  Grab, Gritsan, Grojean, Groom, Gr\"unewald, Gurtu, Gutsche, Haber, Hanhart,
  Hashimoto, Hayato, Hayes, Hebecker, Heinemeyer, Heltsley, Hern\'andez-Rey,
  Hisano, H\"ocker, Holder, Holtkamp, Hyodo, Irwin, Johnson, Kado, Karliner,
  Katz, Klein, Klempt, Kowalewski, Krauss, Kreps, Krusche, Kuyanov, Kwon,
  Lahav, Laiho, Lesgourgues, Liddle, Ligeti, Lin, Lippmann, Liss, Littenberg,
  Lugovsky, Lugovsky, Lusiani, Makida, Maltoni, Mannel, Manohar, Marciano,
  Martin, Masoni, Matthews, Mei\ss{}ner, Milstead, Mitchell, M\"onig, Molaro,
  Moortgat, Moskovic, Murayama, Narain, Nason, Navas, Neubert, Nevski, Nir,
  Olive, Pagan~Griso, Parsons, Patrignani, Peacock, Pennington, Petcov, Petrov,
  Pianori, Piepke, Pomarol, Quadt, Rademacker, Raffelt, Ratcliff, Richardson,
  Ringwald, Roesler, Rolli, Romaniouk, Rosenberg, Rosner, Rybka, Ryutin,
  Sachrajda, Sakai, Salam, Sarkar, Sauli, Schneider, Scholberg, Schwartz,
  Scott, Sharma, Sharpe, Shutt, Silari, Sj\"ostrand, Skands, Skwarnicki, Smith,
  Smoot, Spanier, Spieler, Spiering, Stahl, Stone, Sumiyoshi, Syphers, Terashi,
  Terning, Thoma, Thorne, Tiator, Titov, Tkachenko, T\"ornqvist, Tovey,
  Valencia, Van~de Water, Varelas, Venanzoni, Verde, Vincter, Vogel, Vogt,
  Wakely, Walkowiak, Walter, Wands, Ward, Wascko, Weiglein, Weinberg, Weinberg,
  White, Wiencke, Willocq, Wohl, Womersley, Woody, Workman, Yao, Zeller, Zenin,
  Zhu, Zhu, Zimmermann, Zyla, Anderson, Fuller, Lugovsky, and
  Schaffner]{PhysRevD.98.030001}
Tanabashi, M.; Hagiwara, K.; Hikasa, K.; Nakamura, K.; Sumino, Y.; Takahashi,
  F.; Tanaka, J.; Agashe, K.; Aielli, G.; Amsler, C.; Antonelli, M.; Asner,
  D.M.; Baer, H.; Banerjee, S.; Barnett, R.M.; Basaglia, T.; Bauer, C.W.;
  Beatty, J.J.; Belousov, V.I.; Beringer, J.; Bethke, S.; Bettini, A.; Bichsel,
  H.; Biebel, O.; Black, K.M.; Blucher, E.; Buchmuller, O.; Burkert, V.;
  Bychkov, M.A.; Cahn, R.N.; Carena, M.; Ceccucci, A.; Cerri, A.; Chakraborty,
  D.; Chen, M.C.; Chivukula, R.S.; Cowan, G.; Dahl, O.; D'Ambrosio, G.; Damour,
  T.; de~Florian, D.; de~Gouv\^ea, A.; DeGrand, T.; de~Jong, P.; Dissertori,
  G.; Dobrescu, B.A.; D'Onofrio, M.; Doser, M.; Drees, M.; Dreiner, H.K.;
  Dwyer, D.A.; Eerola, P.; Eidelman, S.; Ellis, J.; Erler, J.; Ezhela, V.V.;
  Fetscher, W.; Fields, B.D.; Firestone, R.; Foster, B.; Freitas, A.;
  Gallagher, H.; Garren, L.; Gerber, H.J.; Gerbier, G.; Gershon, T.; Gershtein,
  Y.; Gherghetta, T.; Godizov, A.A.; Goodman, M.; Grab, C.; Gritsan, A.V.;
  Grojean, C.; Groom, D.E.; Gr\"unewald, M.; Gurtu, A.; Gutsche, T.; Haber,
  H.E.; Hanhart, C.; Hashimoto, S.; Hayato, Y.; Hayes, K.G.; Hebecker, A.;
  Heinemeyer, S.; Heltsley, B.; Hern\'andez-Rey, J.J.; Hisano, J.; H\"ocker,
  A.; Holder, J.; Holtkamp, A.; Hyodo, T.; Irwin, K.D.; Johnson, K.F.; Kado,
  M.; Karliner, M.; Katz, U.F.; Klein, S.R.; Klempt, E.; Kowalewski, R.V.;
  Krauss, F.; Kreps, M.; Krusche, B.; Kuyanov, Y.V.; Kwon, Y.; Lahav, O.;
  Laiho, J.; Lesgourgues, J.; Liddle, A.; Ligeti, Z.; Lin, C.J.; Lippmann, C.;
  Liss, T.M.; Littenberg, L.; Lugovsky, K.S.; Lugovsky, S.B.; Lusiani, A.;
  Makida, Y.; Maltoni, F.; Mannel, T.; Manohar, A.V.; Marciano, W.J.; Martin,
  A.D.; Masoni, A.; Matthews, J.; Mei\ss{}ner, U.G.; Milstead, D.; Mitchell,
  R.E.; M\"onig, K.; Molaro, P.; Moortgat, F.; Moskovic, M.; Murayama, H.;
  Narain, M.; Nason, P.; Navas, S.; Neubert, M.; Nevski, P.; Nir, Y.; Olive,
  K.A.; Pagan~Griso, S.; Parsons, J.; Patrignani, C.; Peacock, J.A.;
  Pennington, M.; Petcov, S.T.; Petrov, V.A.; Pianori, E.; Piepke, A.; Pomarol,
  A.; Quadt, A.; Rademacker, J.; Raffelt, G.; Ratcliff, B.N.; Richardson, P.;
  Ringwald, A.; Roesler, S.; Rolli, S.; Romaniouk, A.; Rosenberg, L.J.; Rosner,
  J.L.; Rybka, G.; Ryutin, R.A.; Sachrajda, C.T.; Sakai, Y.; Salam, G.P.;
  Sarkar, S.; Sauli, F.; Schneider, O.; Scholberg, K.; Schwartz, A.J.; Scott,
  D.; Sharma, V.; Sharpe, S.R.; Shutt, T.; Silari, M.; Sj\"ostrand, T.; Skands,
  P.; Skwarnicki, T.; Smith, J.G.; Smoot, G.F.; Spanier, S.; Spieler, H.;
  Spiering, C.; Stahl, A.; Stone, S.L.; Sumiyoshi, T.; Syphers, M.J.; Terashi,
  K.; Terning, J.; Thoma, U.; Thorne, R.S.; Tiator, L.; Titov, M.; Tkachenko,
  N.P.; T\"ornqvist, N.A.; Tovey, D.R.; Valencia, G.; Van~de Water, R.;
  Varelas, N.; Venanzoni, G.; Verde, L.; Vincter, M.G.; Vogel, P.; Vogt, A.;
  Wakely, S.P.; Walkowiak, W.; Walter, C.W.; Wands, D.; Ward, D.R.; Wascko,
  M.O.; Weiglein, G.; Weinberg, D.H.; Weinberg, E.J.; White, M.; Wiencke, L.R.;
  Willocq, S.; Wohl, C.G.; Womersley, J.; Woody, C.L.; Workman, R.L.; Yao,
  W.M.; Zeller, G.P.; Zenin, O.V.; Zhu, R.Y.; Zhu, S.L.; Zimmermann, F.; Zyla,
  P.A.; Anderson, J.; Fuller, L.; Lugovsky, V.S.; Schaffner, P.
\newblock Review of Particle Physics.
\newblock {\em Phys. Rev. D} {\bf 2018}, {\em 98},~030001.

\bibitem[Aartsen \em{et~al.}(2018)Aartsen et~al.]{Aartsen:2017ibm}
Aartsen, M.G.; others.
\newblock {Neutrino Interferometry for High-Precision Tests of Lorentz Symmetry
  with IceCube}.
\newblock {\em Nature Phys.} {\bf 2018},
  \href{http://xxx.lanl.gov/abs/1709.03434}{{\normalfont
  [arXiv:hep-ex/1709.03434]}}.

\bibitem[Albert \em{et~al.}(2008)Albert et~al.]{Albert:2007qk}
Albert, J.; others.
\newblock {Probing Quantum Gravity using Photons from a flare of the active
  galactic nucleus Markarian 501 Observed by the MAGIC telescope}.
\newblock {\em Phys. Lett.} {\bf 2008}, {\em B668},~253--257,
  \href{http://xxx.lanl.gov/abs/0708.2889}{{\normalfont
  [arXiv:astro-ph/0708.2889]}}.

\bibitem[Martinez and Errando(2009)]{Martinez:2008ki}
Martinez, M.; Errando, M.
\newblock {A new approach to study energy-dependent arrival delays on photons
  from astrophysical sources}.
\newblock {\em Astropart. Phys.} {\bf 2009}, {\em 31},~226--232,
  \href{http://xxx.lanl.gov/abs/0803.2120}{{\normalfont
  [arXiv:astro-ph/0803.2120]}}.

\bibitem[Abdo \em{et~al.}(2009)Abdo et~al.]{Ackermann:2009aa}
Abdo, A.A.; others.
\newblock {A limit on the variation of the speed of light arising from quantum
  gravity effects}.
\newblock {\em Nature} {\bf 2009}, {\em 462},~331--334,
  \href{http://xxx.lanl.gov/abs/0908.1832}{{\normalfont
  [arXiv:astro-ph.HE/0908.1832]}}.

\bibitem[Abramowski \em{et~al.}(2011)Abramowski et~al.]{HESS:2011aa}
Abramowski, A.; others.
\newblock {Search for Lorentz Invariance breaking with a likelihood fit of the
  PKS 2155-304 Flare Data Taken on MJD 53944}.
\newblock {\em Astropart. Phys.} {\bf 2011}, {\em 34},~738--747,
  \href{http://xxx.lanl.gov/abs/1101.3650}{{\normalfont
  [arXiv:astro-ph.HE/1101.3650]}}.

\bibitem[Nemiroff \em{et~al.}(2012)Nemiroff, Connolly, Holmes, and
  Kostinski]{Nemiroff:2011fk}
Nemiroff, R.J.; Connolly, R.; Holmes, J.; Kostinski, A.B.
\newblock {Bounds on Spectral Dispersion from Fermi-detected Gamma Ray Bursts}.
\newblock {\em Phys. Rev. Lett.} {\bf 2012}, {\em 108},~231103,
  \href{http://xxx.lanl.gov/abs/1109.5191}{{\normalfont
  [arXiv:astro-ph.CO/1109.5191]}}.

\bibitem[Vasileiou \em{et~al.}(2013)Vasileiou, Jacholkowska, Piron, Bolmont,
  Couturier, Granot, Stecker, Cohen-Tanugi, and Longo]{Vasileiou:2013vra}
Vasileiou, V.; Jacholkowska, A.; Piron, F.; Bolmont, J.; Couturier, C.; Granot,
  J.; Stecker, F.W.; Cohen-Tanugi, J.; Longo, F.
\newblock {Constraints on Lorentz Invariance Violation from Fermi-Large Area
  Telescope Observations of Gamma-Ray Bursts}.
\newblock {\em Phys. Rev.} {\bf 2013}, {\em D87},~122001,
  \href{http://xxx.lanl.gov/abs/1305.3463}{{\normalfont
  [arXiv:astro-ph.HE/1305.3463]}}.

\bibitem[Vasileiou \em{et~al.}(2015)Vasileiou, Granot, Piran, and
  Amelino-Camelia]{Vasileiou:2015wja}
Vasileiou, V.; Granot, J.; Piran, T.; Amelino-Camelia, G.
\newblock {A Planck-scale limit on spacetime fuzziness and stochastic Lorentz
  invariance violation}.
\newblock {\em Nature Phys.} {\bf 2015}, {\em 11},~344--346.

\bibitem[Carmona \em{et~al.}(2018)Carmona, Cortes, and
  Relancio]{Carmona:2017oit}
Carmona, J.M.; Cortes, J.L.; Relancio, J.J.
\newblock {Does a modification of special relativity imply energy dependent
  photon time delays?}
\newblock {\em Class. Quant. Grav.} {\bf 2018}, {\em 35},~025014,
  \href{http://xxx.lanl.gov/abs/1702.03669}{{\normalfont
  [arXiv:hep-th/1702.03669]}}.

\bibitem[Carmona \em{et~al.}(2012)Carmona, Cort\'es, and Mercati]{Carmona2012}
Carmona, J.M.; Cort\'es, J.L.; Mercati, F.
\newblock {Relativistic kinematics beyond Special Relativity}.
\newblock {\em Phys.Rev.} {\bf 2012}, {\em D86},~084032,
  \href{http://xxx.lanl.gov/abs/1206.5961}{{\normalfont
  [arXiv:hep-th/1206.5961]}}.

\bibitem[Carmona \em{et~al.}(2016)Carmona, Cortes, and Relancio]{Carmona2016b}
Carmona, J.M.; Cortes, J.L.; Relancio, J.J.
\newblock {Beyond Special Relativity at second order}.
\newblock {\em Phys. Rev.} {\bf 2016}, {\em D94},~084008,
  \href{http://xxx.lanl.gov/abs/1609.01347}{{\normalfont
  [arXiv:hep-th/1609.01347]}}.

\bibitem[Majid and Ruegg(1994)]{Majid1994}
Majid, S.; Ruegg, H.
\newblock {Bicrossproduct structure of kappa Poincare group and noncommutative
  geometry}.
\newblock {\em Phys. Lett.} {\bf 1994}, {\em B334},~348--354,
  \href{http://xxx.lanl.gov/abs/hep-th/9405107}{{\normalfont
  [arXiv:hep-th/hep-th/9405107]}}.

\bibitem[Carmona \em{et~al.}(2018)Carmona, Cortes, and
  Relancio]{Carmona:2017cry}
Carmona, J.M.; Cortes, J.L.; Relancio, J.J.
\newblock {Spacetime from locality of interactions in deformations of special
  relativity: The example of $\kappa$-Poincar\'e Hopf algebra}.
\newblock {\em Phys. Rev.} {\bf 2018}, {\em D97},~064025,
  \href{http://xxx.lanl.gov/abs/1711.08403}{{\normalfont
  [arXiv:hep-th/1711.08403]}}.

\bibitem[Borowiec and Pachol(2010)]{Borowiec2010}
Borowiec, A.; Pachol, A.
\newblock {Classical basis for kappa-Poincare algebra and doubly special
  relativity theories}.
\newblock {\em J. Phys.} {\bf 2010}, {\em A43},~045203,
  \href{http://xxx.lanl.gov/abs/0903.5251}{{\normalfont
  [arXiv:hep-th/0903.5251]}}.

\bibitem[Kowalski-Glikman and Nowak(2002)]{Kowalski-Glikman2002}
Kowalski-Glikman, J.; Nowak, S.
\newblock {Doubly special relativity theories as different bases of kappa
  Poincare algebra}.
\newblock {\em Phys. Lett.} {\bf 2002}, {\em B539},~126--132,
  \href{http://xxx.lanl.gov/abs/hep-th/0203040}{{\normalfont
  [arXiv:hep-th/hep-th/0203040]}}.

\bibitem[Battisti and Meljanac(2010)]{Battisti:2010sr}
Battisti, M.V.; Meljanac, S.
\newblock {Scalar Field Theory on Non-commutative Snyder Space-Time}.
\newblock {\em Phys. Rev.} {\bf 2010}, {\em D82},~024028,
  \href{http://xxx.lanl.gov/abs/1003.2108}{{\normalfont
  [arXiv:hep-th/1003.2108]}}.

\bibitem[Majid(1995)]{Majid:1995qg}
Majid, S.
\newblock {\em Foundations of Quantum Group Theory}; Cambridge University
  Press,  1995.

\bibitem[Atsue and Oyewande(2015)]{Atsue2015}
Atsue, T.; Oyewande, E.
\newblock Investigation of the Effect of Interference of Photon and Z0 Boson
  Exchanges on the Energy Dependence of Muon Pair Production in Electron
  Positron Annihilation.
\newblock {\em International Journal of High Energy Physics} {\bf 2015}, {\em
  2},~56--60.

\bibitem[Mironov and Morozov(2017)]{Mironov:2017gja}
Mironov, A.; Morozov, A.
\newblock {Check-Operators and Quantum Spectral Curves}.
\newblock {\em SIGMA} {\bf 2017}, {\em 13},~047,
  \href{http://xxx.lanl.gov/abs/1701.03057}{{\normalfont
  [arXiv:hep-th/1701.03057]}}.

\end{thebibliography}
\end{document}